\documentclass[12pt,preprint]{aastex}

\newcommand{\is}[2]{\ensuremath{^{#1}\mathrm{#2}}}
\newcommand{\mhc}{\ensuremath{M_5^{\mathrm{HC}}}}

\begin{document}

\title{Neutrino-Accelerated Hot Hydrogen Burning}

\author{Chad T. Kishimoto and George M. Fuller}
\affil{Department of Physics, University of California, San Diego, La Jolla, CA 92093-0319}

\begin{abstract}

We examine the effects of significant electron anti-neutrino fluxes on hydrogen burning.  Specifically, we find that the bottleneck weak nuclear reactions in the traditional pp-chain and the hot CNO cycle can be accelerated by anti-neutrino capture, increasing the energy generation rate.  We also discuss how anti-neutrino capture reactions can alter the conditions for break out into the $rp$-process.  We speculate on the impact of these considerations for the evolution and dynamics of collapsing very- and super- massive compact objects.

\end{abstract}

\section{Introduction}

Hydrogen burning involves the conversion of four protons into an alpha particle, two positrons, neutrinos and photons.  The principal bottleneck involved in this process is the weak interaction conversion of protons into neutrons.  For decades the primary mechanisms of hydrogen burning have been an astronomical staple.  \citet{bet38} first elucidated the proton-proton chain (pp-chain) where the weak conversion is accomplished by two protons interacting to become a deuteron, $p ( p, \nu_e e^+ ) d$.  \citet{vwe37, vwe38} and \citet{bet39} independently described the CNO cycle, where carbon is used as a catalyst in hydrogen burning, and the weak conversion of protons to neutrons occurs through the positron decay of isotopes of oxygen with half lives of about 100 seconds.

A large flux of electron anti-neutrinos ($\bar\nu_e$) could alter the hydrogen burning paradigm.  Anti-neutrino capture could perform the necessary conversion of protons to neutrons.  The $\bar\nu_e$-capture cross sections of relevance are very small, but depend strongly on neutrino energy.  The smallness of these cross sections allows energetic neutrinos to escape from deep within a compact object, where the temperature and other energy scales are high, and freely stream to where hydrogen burning is occurring.  Nevertheless, if $\bar\nu_e$-capture is to have a significant effect on hot hydrogen burning, a truly prodigious flux ($\phi_{\bar\nu_e} \gtrsim 10^{40} ~\mbox{cm}^{-2} ~\mbox{s}^{-1}$) and large neutrino energy ($\langle E_{\bar\nu_e} \rangle \gtrsim \mbox{a few MeV}$) would be necessary.  It should be kept in mind, however, that to affect hydrogen burning, the $\bar\nu_e$-capture rates need only be comparable to the corresponding positron decay rates.

The difficulty would be to find an environment capable of producing these fluxes of neutrinos, yet quiescent enough that simple hydrogen burning could be relevant, and the products of such burning could be ejected into space.  High entropy electron-positron plasmas are efficient engines for the production of neutrinos and anti-neutrinos of all flavors.  Possible environments that may merit future investigations into the effects of anti-neutrino capture on hydrogen burning include high mass accretion disks and collapsing very- and super- massive objects.

In this paper, we investigate the effects of a prodigious neutrino flux on hot hydrogen burning.  In section \ref{burnmech} we point out the effects of anti-neutrino capture on the rate limiting steps in both the pp-chain and the $\beta$-limited CNO cycle, and its implications for the energy generation rates.  In section \ref{consequences}, we examine the consequences for the $rp$-process and energy generation mechanisms.  In section \ref{smo}, we consider the case of a supermassive star collapsing on the general relativistic Feynman-Chandrasekhar instability, and the effects of its internal neutrino production on hydrogen burning in its envelope.  We give conclusions in section \ref{conclusions}.

\section{Neutrino-Induced Hydrogen Burning Mechanisms} \label{burnmech}

The rate limiting step in the pp-chain is the weak interaction conversion of two protons into a deuteron, a positron, and an electron neutrino.  A significant flux of electron anti-neutrinos allows an alternate mechanism to be favored, where anti-neutrino capture on a proton creates a neutron and a positron \citetext{$\bar\nu_e + p \rightarrow n + e^+$ has been considered in supermassive objects by \citet{woo77} and \citet{ful97}}.  This step would be followed by a fast radiative proton capture to form a deuteron.  Comparing the two reaction rates: $p( p, \nu_e e^+) d$ vs. $p( \bar\nu_e, e^+ ) n( p, \gamma ) d$, we find that for the prodigious anti-neutrino fluxes discussed in the introduction ($\phi_{\bar\nu_e} \gtrsim 10^{40} ~\mbox{cm}^{-2} ~\mbox{s}^{-1}$, $\langle E_{\bar\nu_e} \rangle \gtrsim \mbox{a few MeV}$) the anti-neutrino capture path is significantly faster in relevant astrophysical environments.  This provides not only a new reaction path for hydrogen burning, but increases the energy generation rate by several orders of magnitude.

\begin{figure}
\epsscale{.75}
\plotone{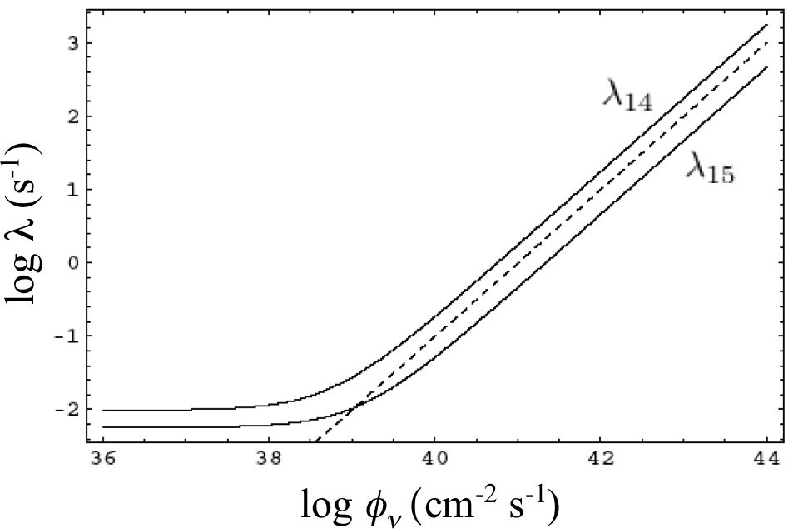}
\caption{Key weak decay rates as a function of electron anti-neutrino flux, assuming a Fermi-Dirac $\bar\nu_e$-energy spectrum with zero chemical potential and average $\bar\nu_e$-energy $\langle E_{\bar\nu_e} \rangle = 10 ~\mbox{MeV}$.  The solid lines, labeled $\lambda_{14}$ and $\lambda_{15}$ are the total weak decay rates of \is{14}{O} and \is{15}{O}, respectively, through positron decay and anti-neutrino capture.  The dashed line is the rate of conversion of a proton to a neutron via anti-neutrino capture alone.\label{figure1}}
\end{figure}

The $\beta$-limited CNO cycle, or hot CNO cycle, proceeds at a rate dictated by the positron decay of \is{14}{O} and \is{15}{O}, with half lives of $71\,{\rm s}$ and $122\,{\rm s}$, respectively.  \citetext{See for example \citealt{hoy65} and \citealt{aud73}.}  These decays likewise could be augmented by electron anti-neutrino capture, $\is{14}{O}(\bar\nu_e, e^+)\is{14}{N}$ and $\is{15}{O}(\bar\nu_e, e^+)\is{15}{N}$.  Figure \ref{figure1} shows the acceleration of the relevant weak rates as a function of total electron anti-neutrino flux for an assumed Fermi-Dirac $\bar\nu_e$-energy spectrum with average $\bar\nu_e$-energy $\langle E_{\bar\nu_e} \rangle = 10 ~\mbox{MeV}$ and zero chemical potential.  The flux at which anti-neutrino capture becomes important scales appoximately as $\langle E_{\bar\nu_e} \rangle^{-2}$.  For a large enough flux ($\phi_{\bar\nu_e} \gtrsim 10^{39} ~\mbox{cm}^{-2} ~\mbox{s}^{-1}$ in the case of Figure \ref{figure1}), the reaction rates are proportional to the incident flux of electron anti-neutrinos.  Our weak rate calculations are described in Appendix \ref{ratecalc}.

\begin{figure}
\epsscale{.75}
\plotone{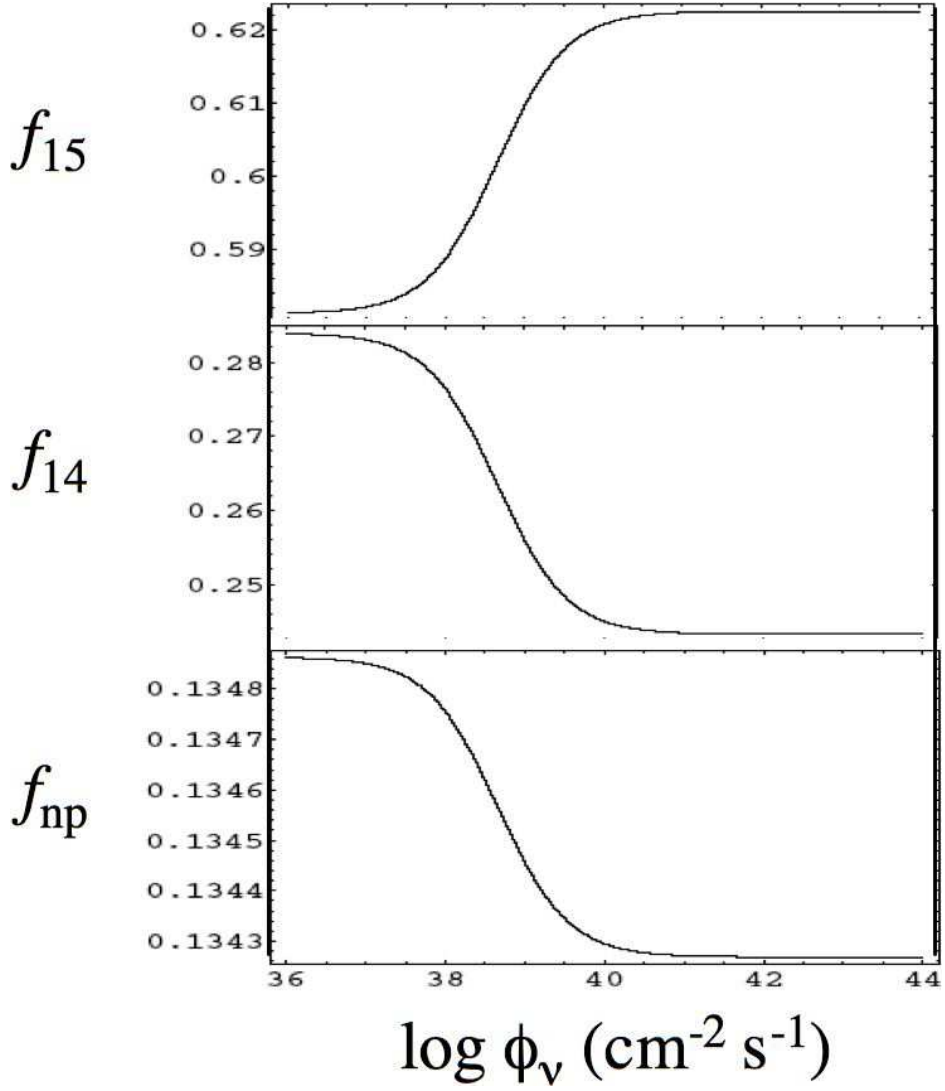}
\caption{Fraction of neutron captures on \is{15}{O} ($f_{15}$), \is{14}{O} ($f_{14}$) and $p$ ($f_{np}$), assuming a Fermi-Dirac $\bar\nu_e$-energy spectrum with zero chemical potential, average $\bar\nu_e$-energy $\langle E_{\bar\nu_e} \rangle = 10 ~\mbox{MeV}$ and $X / Z' = 100$, where $X$ is the hydrogen mass fraction and $Z'$ is the mass fraction in carbon, nitrogen and oxygen isotopes.\label{figure4}}
\end{figure}

Additionally, the CNO cycle is accelerated by the presence of free neutrons.  The strong interaction reactions $\is{15}{O} (n, p) \is{15}{N}$ and $\is{14}{O} (n, p) \is{14}{N}$ have a significantly larger cross section than the electromagnetic reaction $n (p, \gamma) d$.  As a result, neutrons are diverted from the modified pp-chain into the CNO cycle.  Figure \ref{figure4} shows how the neutrons created by $p (\bar\nu_e, e^+) n$ are distributed between the competing reactions $\is{15}{O} (n, p) \is{15}{N}$, $\is{14}{O} (n, p) \is{14}{N}$ and $n (p, \gamma) d$.  Notice that for an assumed Fermi-Dirac $\bar\nu_e$-energy spectrum with $\langle E_{\bar\nu_e} \rangle = 10 ~\mbox{MeV}$ and zero chemical potential, the ratio of neutron captures on \is{15}{O} to \is{14}{O} to $p$ is approximately $4.5 : 2 : 1$ for a large range of $\bar\nu_e$-fluxes.

\begin{figure}
\epsscale{.75}
\plotone{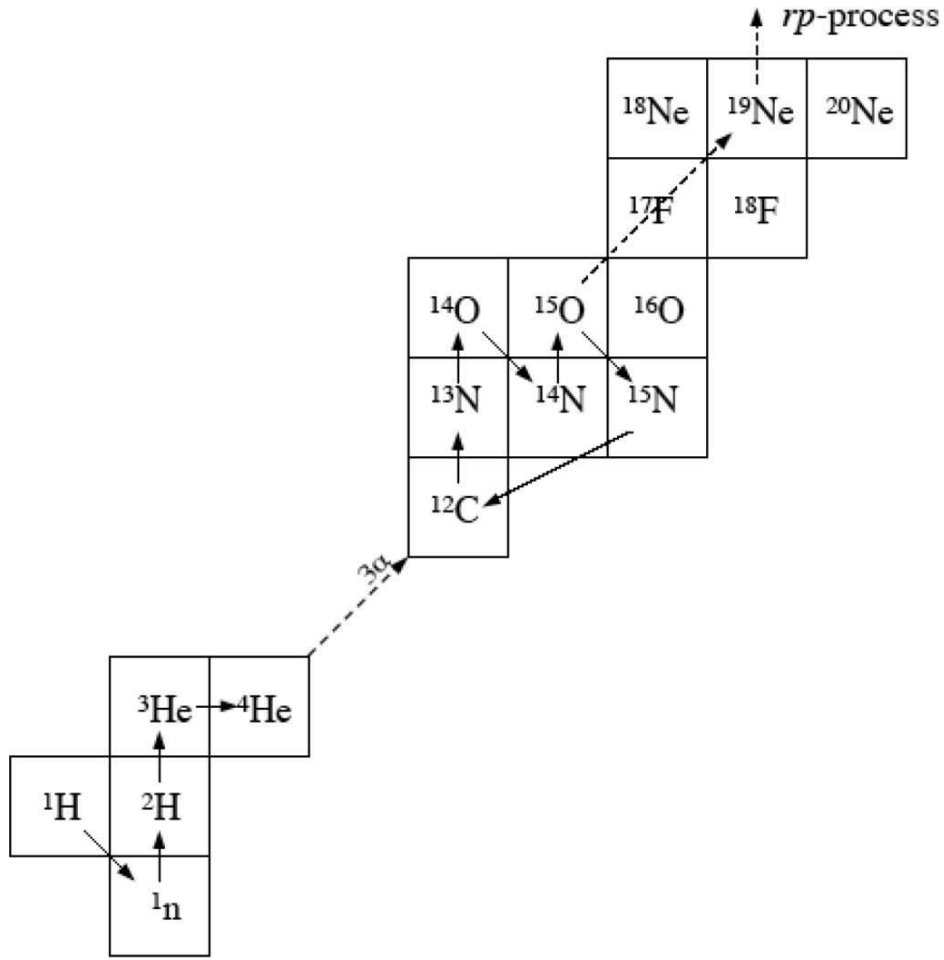}
\caption{Qualitative picture of hydrogen burning under the influence of a prodigious electron anti-neutrino flux.  Shown are a modified pp-chain, $\beta$-limited CNO cycle, and the less significant triple-$\alpha$ reaction and break-out to the $rp$-process.\label{figure2}}
\end{figure}

Figure \ref{figure2} illustrates the most significant reaction flow paths involved in hydrogen burning when a significant $\bar\nu_e$-flux is present.  The pp-chain is modified as anti-neutrino capture allows the circumvention of the slow $p ( p, \nu_e e^+ ) d$ reaction.  Also included are the triple-alpha process, which would provide a path between the pp-chain and CNO cycle, and the break out into the $rp$-process via $\is{15}{O}( \alpha, \gamma )\is{19}{Ne}( p, \gamma )\is{20}{Na}$. \citep{wal81}

\section{Side Effects} \label{consequences}

A large flux of electron anti-neutrinos certainly accelerates the weak rates that provide the bottleneck in hot hydrogen burning.  However, since this flux also increases the rates of other positron decays, a number of side effects are possible.

A principal mechanism for break out into the $rp$-process involves the reaction path  $\is{15}{O} ( \alpha, \gamma ) \is{19}{Ne}( p, \gamma ) \is{20}{Na}$.  The criteria for break out into the $rp$-process can be found in the competition between proton capture on \is{19}{Ne}, and the decay of \is{19}{Ne} through positron emission and now, anti-neutrino capture.  Thus for densities and temperatures that satisfy the inequality
\begin{equation}
\label{rpbreakout}
\rho X \lambda_{p \gamma} (\is{19}{Ne}) > \lambda_{w} (\is{19}{Ne}),
\end{equation}
break out into the $rp$-process will occur \citep{wal81}.  Here the density $\rho$ is in $\mbox{g} \,\mbox{cm}^{-3}$, $X$ is the hydrogen mass fraction, $\lambda_{w} (\is{19}{Ne})$ is the total weak decay rate of \is{19}{Ne} (positron emission and $\bar\nu_e$-capture), and $\lambda_{p\gamma} = N_A \langle \sigma v \rangle_{p\gamma}$, where $N_A$ is Avagadro's number and the thermally averaged product of cross section and speed is taken from \citet{cf88}.

\begin{figure}
\epsscale{.75}
\plotone{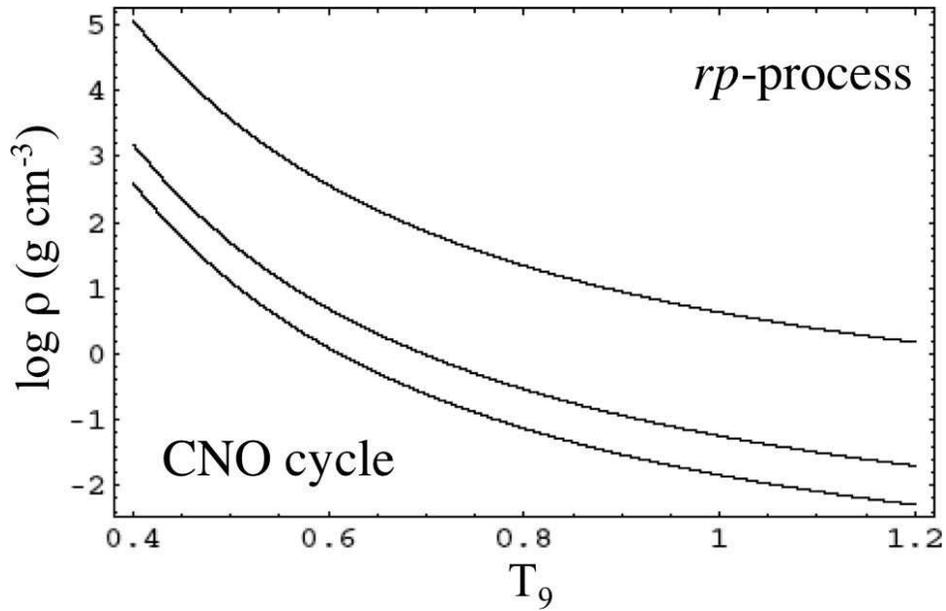}
\caption{Conditions for break-out into the $rp$-process, assuming a Fermi-Dirac $\bar\nu_e$-energy spectrum, zero chemical potential, and $\langle E_{\bar\nu_e} \rangle = 10 ~\mbox{MeV}$.  The plotted contours are for, in ascending order, $\phi_{\bar\nu_e} = 10^{38,40,42} ~\mbox{cm}^{-2} ~\mbox{s}^{-1}$.  Zero neutrino flux is indistinguishable from $\phi_{\bar\nu_e} = 10^{38} ~\mbox{cm}^{-2} ~\mbox{s}^{-1}$.\label{figure3}}
\end{figure}

Including a large flux of electron anti-neutrinos would result in higher weak decay rates ($\lambda_{w}$).  This increase in the right hand side of equation (\ref{rpbreakout}) would require an increase in temperature (increasing $\lambda_{p\gamma}(\is{19}{Ne})$) for a given density at which break out into the $rp$-process would occur.  Figure \ref{figure3} shows the effects of an electron anti-neutrino flux on the conditions necessary for break out into the $rp$-process.

The pp-chain is the dominant process of energy generation in the sun, while the CNO cycle is dominant in stars that are more massive.  However, with a large flux of electron anti-neutrinos, these processes become independent of temperature so long as the temperature is high enough to guarantee that proton capture remains comparatively fast.  A high flux of electron anti-neutrinos allows the pp-chain to compete favorably with the CNO cycle.  For example, \citet{pru05} have studied nucleosynthesis in supernova winds where hydrogen ``burning'' is completely dominated by $\nu_e$ and $\bar\nu_e$ capture on free nucleons \citep[see also][]{qian93}.

\begin{figure}
\epsscale{.75}
\plotone{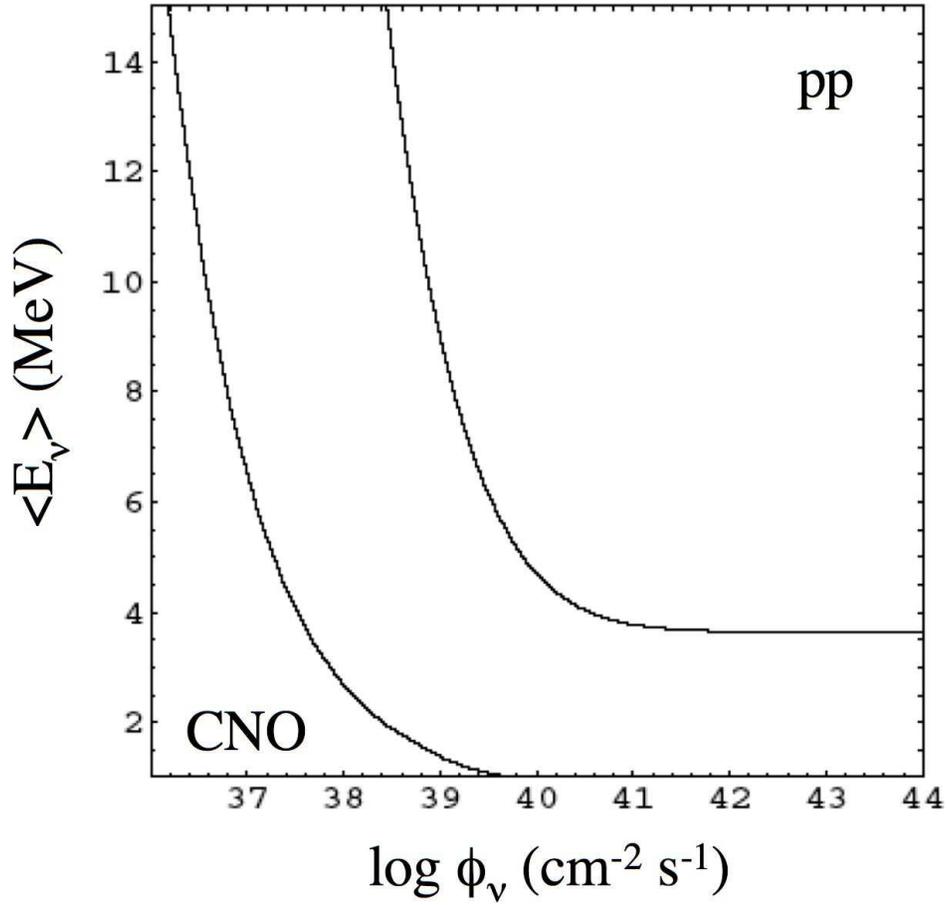}
\caption{Comparison of energy generation rates for the pp-chain and CNO cycle at a temperature of $5 \times 10^8 ~\mbox{K}$.  The contours are where the energy generation rates are equal for the two hydrogen burning mechanisms for values of $X / Z' = 10, 100$ (from right to left).  Larger $\bar\nu_e$-fluxes and average energies favors the pp-chain, while smaller fluxes and average energies favor the CNO cycle.\label{figure5}}
\end{figure}

The scarcity of \is{15}{O} in comparison to free protons means that for large anti-neutrino fluxes and average energies, the pp-chain is the dominant mechanism in hydrogen burning at temperatures that the CNO cycle would traditionally dominate.  Figure \ref{figure5} shows a comparison between the energy generation rates of the pp-chain and the CNO cycle for $X / Z' = 10$ and $100$, where $X$ is the hydrogen mass fraction and $Z'$ is the mass fraction in carbon, nitrogen, and oxygen isotopes.  For large anti-neutrino fluxes and average energies, the pp-chain is the dominant energy generation mechanism, while for low fluxes and average energies the CNO cycle takes over.

\section{Example:  Supermassive Stars} \label{smo}

Now we consider the case of a supermassive star, a star so massive that it collapses on the general relativistic Feynman-Chandrasekhar instability ($M \gtrsim 5 \times 10^4 M_\odot$) \cite[hereafter FWW]{fww86}.  If such objects did exist, for example in the early universe, their homologous cores would emit copious fluxes of neutrinos and anti-neutrinos of all flavors during their collapse \citep{shi98}.  Shi and Fuller examined the collapsing core of a supermassive star, calculating the luminosity and energy spectrum of neutrinos emitted.  The total neutrino luminosity was found to be
\begin{equation}
L_\nu \approx 2.8 \times 10^{57} \left( \mhc \right)^{-1.5} ~\mbox{erg} ~\mbox{s}^{-1},
\end{equation}
where \mhc{} is the mass of the homologous core in units of $10^5 M_\odot$.  Additionally they found the energy spectrum of neutrinos of all flavors to fit remarkably well to a Fermi-Dirac spectrum with a higher temperature than the central plasma temperature ($T_\nu \approx 1.6 T$) and a degeneracy parameter (chemical potential divided by temperature) $\eta_\nu \approx 2$.

We can check the effect of this flux of neutrinos an antineutrinos on the nuclear physics in the gas in the envelope of the star.  As a point of reference, we choose a radius of 100 Schwarzschild radii ($r = 3 \times 10^{12} \mhc ~\mbox{cm}$) where the gravitational binding energy of a nucleon is approximately equal to the nuclear energy liberated in these reactions, so there is a chance that any new nuclear physics that occurs as a result of $\bar\nu_e$-capture could be relevant.  By ``relevant'' we mean that it is conceivable that material from this location could avoid being swallowed by the black hole forming in the core.  Only a detailed simulation with general relativistic hydrodynamics could reveal whether or not material affected by $\bar\nu_e$-capture is ever ejected into space.  We are now free to repeat the analyses done above, but with one free parameter, the mass of the collapsing homologous core.

\begin{figure}
\epsscale{.75}
\plotone{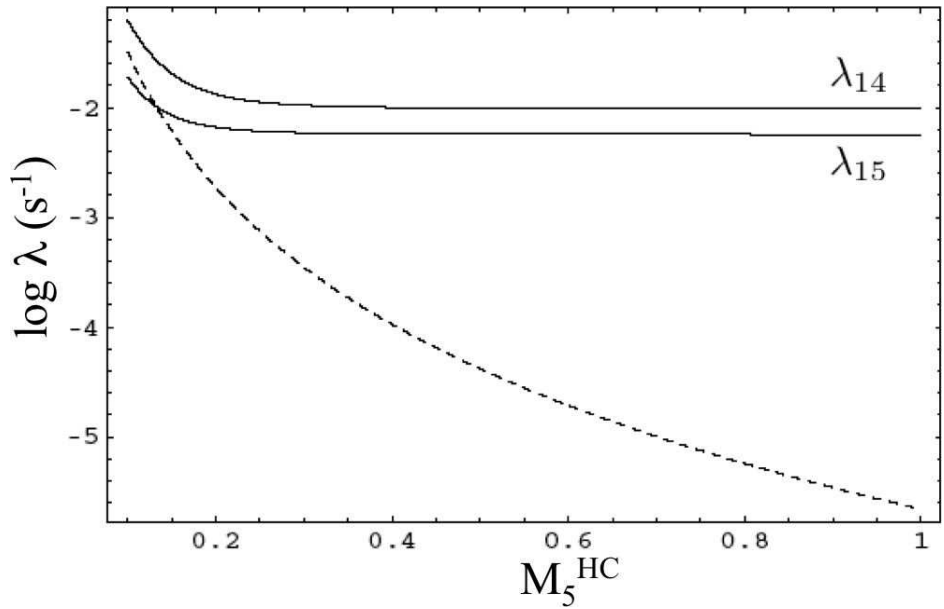}
\caption{Rates of relevant weak decays at a distance of 100 Schwartzschild radii from the center of a supermassive star with homologous core mass $\mhc \times 10^5 M_\odot$.  As in Figure \ref{figure1}, the solid lines are the total rates of decay of \is{14,15}{O}, and the dashed line is the proton to neutron conversion rate through anti-neutrino capture alone.\label{figure6}}
\end{figure}

Figure \ref{figure6} shows the acceleration of the relevant weak rates as a function of homologous core mass.  The anti-neutrino capture rate is proportional to $(\mhc)^{-4}$.  We see that the effects of anti-neutrino capture on the decay rates of \is{14}{O} and \is{15}{O} become insignificant for large homologous core masses ($\mhc \gtrsim 0.4$).  If supermassive stars ever formed, it is conceivable that they were in the first generation of stars with primordial initial abundances.  In this case the CNO cycle would be negligible, at least initially \citetext{FWW}.  However, even in this case, the energy generation rate of the pp-chain would be boosted by several orders of magnitude.  It would be interesting to see if this added energy source would have a discernible effect on the eventual fate of a collapsing supermassive star.

Including a hydrogen burning phase in the final stages of the collapse of a supermassive star may affect the eventual fate of its baryons.  Hydrodynamic, post-Newtonian calculations done in FWW show that initially metal-free supermassive stars will collapse to black holes.  \citet{shi02} use a fully relativistic simulation in axial symmetry to deduce that the supermassive star collapses to a black hole surrounded by some remaining gas in an ambient disk.  

The principal nucleosynthetic issue is whether any material that had experienced $\bar\nu_e$ capture-affected hydrogen burning escapes being incorporated into a black hole.  Of course, there is the prior issue of whether material at the relatively low temperatures and densities which characterize hydrogen burning ever experiences high $\bar\nu_e$-fluxes.  Both issues are related:  to see nucleosynthesis products of $\bar\nu_e$ capture-affected hydrogen burning, the material must be ejected before the point at which nuclear burning proceeds past simple hydrogen burning and approaches or attains nuclear statistical equilibrium (NSE).  We are skeptical that these conditions can be met.  Fully relativistic numerical simulations could settle these issues.

Obviously, the NSE nucleosynthetic yield is uninteresting in the context of this paper.  However, a mass shedding scenario could be conducive to conditions that favor hydrogen burning and the $rp$-process.    Speeding up weak decays could affect the relative abundances of the $rp$-process elements.  A simulation that follows these species and their chemical reactions would be necessary to address this issue.

\section{Conclusions} \label{conclusions}

In this paper we have examined the effects of a prodigious flux of electron anti-neutrinos on hydrogen burning.  We have found that the traditional positron decay bottlenecks in hydrogen burning can be removed and replaced by much faster $\bar\nu_e$-capture reactions under some conditions.  This would result in an increase of several orders of magnitude in the energy generation rate over what would be expected without such a flux. 

Additionally, the $\bar\nu_e$-flux would alter the conditions necessary for break-out into the $rp$-process, increasing the temperature necessary to do so at a given density.  If conditions allow the break-out into the $rp$-process, we could expect an acceleration of the flow toward the iron-peak facilitated by and accelerated by $\bar\nu_e$-capture.

When applied to the neutrino flux emitted in the final stages of the collapse of a supermassive star, interesting changes from current simulations may occur on the lower end of the supermassive star mass spectrum.  Whether or not these effects are relevant, remains an open question that can only be answered by simulations that are able to include hydrogen burning during the final collapse of the star.  

Important issues that remain open include finding an astrophysical environment where the effects discussed here could take place.  Accretion disks surrounding black holes may provide a combination of high accretion rates and hot, high entropy disks which could produce the necessary fluxes of electron anti-neutrinos.  (See for example \citet{sur05} for a discussion of neutrino emission in lower mass accretion disks.) Supermassive stars may exhibit this effect, though there is uncertainty related to whether or not these objects ever existed.  Computer simulations would be useful to determine any changes in the expected nucleosynthetic yield, and the effects of their possible distribution into the surrounding IGM.

\acknowledgements

We would like to thank S. E. Woosley, Y.-Z. Qian and A. Heger for useful discussions.  This work was supported in part by NSF grant PHY-04-00359 and the TSI collaboration's DOE SciDAC grant at UCSD.  C.T.K. would like to acknowledge a fellowship from the ARCS Foundation, Inc.

\appendix

\section{Calculation of Weak Rates} \label{ratecalc}

In this work we calculate the $\bar\nu_e$-capture rates in the manner described in \citet{ffn1,ffn2} \citetext{FFN I, II} and \citet{fm85}.  We employ measured discrete states only.

Our $\is{14}{O} (\bar\nu_e, e^+) \is{14}{N}$ rate calculation includes only the \is{14}{O} ground state (spin and parity $J^\pi = 0^+$) and the measured weak branches to the \is{14}{N} ground state ($J^\pi = 0^+$, $\log_{10} ft = 7.3$), first excited state ($J^\pi = 1^+$, $\log_{10} ft = 3.5$), and second excited state ($J^\pi = 1^+$, $\log_{10} ft = 3.1$).  Contributions to the stellar rate from thermal excitation of parent states are small here as a result of the high first excited state excitation energy ($5.17 ~\mbox{MeV}$) and the temperatures of interest.  Likewise, branches to higher excited states in \is{14}{N} are not significant.  A possible exception is the first isobaric analog state in \is{14}{N} ($J^\pi = 0^+$) at excitation energy of $8.62 ~\mbox{MeV}$.  This branch will have a large matrix element but will be $Q$-value-hindered relative to the $0^+ \rightarrow 0^+$ ground state to first excited state, pure Fermi branch.

Our $\is{15}{O} (\bar\nu_e, e^+) \is{15}{N}$ rate calculation includes only the ground state ($J^\pi = 1/2^-$) branch.  This channel has a large matrix element, corresponding to $\log_{10} ft = 3.6$.  Branches to \is{15}{N} excited states will not be significant.  \is{15}{N} states below $9.15 ~\mbox{MeV}$ excitation energy have positive parity and the branches from the \is{15}{O} ground state to them will be forbidden.  We note, however, that \is{15}{O} and \is{15}{N} are isospin mirrors.  This can be a significant fact for stellar weak interaction rates, as it implies large Fermi and Gamow-Teller matrix elements coupling each parent state with its daughter isobaric analog state \citetext{FFN I, II}.  Thermal excitation of \is{15}{O} excited states would open weak branches to corresponding isobaric analog states in \is{15}{N}.  This is not likely at the temperatures of interest because the first excited state of \is{15}{O} is at about $5.2 ~\mbox{MeV}$ excitation.

Our calculation of the $\is{19}{Ne}( \bar\nu_e, e^+ ) \is{20}{Na}$ rate includes only the ground state of \is{19}{Ne} ($J^\pi = 1/2^+$) and branches to the ground ($J^\pi = 1/2^+$) and third excited state ($J^\pi = 3/2^+$) of \is{19}{F}.  The first of these branches, with $\log_{10} ft = 3.2$, dominates the rate.  We note, however, that \is{19}{Ne} and \is{19}{F} are isospin mirrors.  Since temperatures are high near CNO cycle breakout, thermal excitation of the first ($J^\pi = 5/2^+$) and second ($J^\pi = 1/2^-$) excited states can be expected to carry a fraction of the total weak rate.  However, on the assumption that the matrix elements for these branches are identical to that for the ground-to-ground transitions, inclusion of these branches makes little difference ($< 1\%$) for the rates and our conclusions.

\clearpage

 \end{document}